\documentstyle[12pt,preprint,aps,epsf,amsfonts,floats]{revtex} 
\tighten 
\newcommand{\be}{\begin{equation}}
\newcommand{\ee}{\end{equation}}
\newcommand{\bea}{\begin{eqnarray}}
\newcommand{\eea}{\end{eqnarray}}
\newcommand{\beaa}{\begin{eqnarray}}
\newcommand{\eeaa}{\end{eqnarray}}
\newcommand{\ba}{\begin{array}}
\newcommand{\ea}{\end{array}}
\newcommand{\bit}{\begin{itemize}}
\newcommand{\eit}{\end{itemize}}
\newcommand{\ben}{\begin{enumerate}}
\newcommand{\een}{\end{enumerate}}

\def\lab{\label}
\def\lan{\langle}

\def\lf{\left}
\def\lrar{\leftrightarrow}

\def\noi{\noindent}
\def\non{\nonumber}

\def\pa{\partial}
\def\ran{\rangle}
\def\rar{\rightarrow}

\def\ri{\right}
\def\ti{\tilde}

\def\de{\delta}
\def\De{\Delta}

\def\te{\theta}

\def\la{\lambda}
\def\La{\Lambda}

\def\Om{\Omega}
\begin{document}
\title{Quantum Field Theory of Topological Defects \\[2mm] 
as Inhomogeneous Condensates}

\author{{\bf Massimo Blasone$^\sharp$} and {\bf Petr
Jizba$^\flat$}\thanks{e-mail: M.Blasone@ic.ac.uk,
P.Jizba@damtp.cam.ac.uk. 
To appear in the Proceedings of the 
37th International School of Subnuclear 
Physics, Erice, 29 Aug. - 7 Sep. 1999.
}\\[2mm]}

\address{${}^\sharp$ Blackett Laboratory, Imperial College,
 London SW7 2BZ, U.K.  
\\ and  Dipartimento di Fisica dell'Universit\`a di  Salerno \\
 and INFN, I-84100 Salerno, Italy \\[2mm] 
${}^\flat$ DAMTP, Silver Street, Cambridge, U.K.}

\maketitle 
\vspace{3mm}

\centerline{\bf Contents:}
\ben

\item
Introduction
\item
The Haag expansion in Closed--Time--Path formalism
\item
Kinks in two--dimensional $\la \psi^4$ theory
\item
Vortices in four--dimensional $\la \psi^4$ theory
\item
Conclusions
\een

\vspace{1.5cm}

%%%%%%%%%%%%%%%%%%%%%%%%%%%%%%%%%%%%%%%%%%%%%%%%%%%%%%%%%%%%%%%%%

\noi{\bf  1. Introduction}

\vspace{0.2cm}

Topological defects play an important r\^ole in many physical systems
ranging from cosmology to condensed matter. Thus they link 
apparently unrelated areas characterized by very
different energy and time scales.

The issue of spontaneous 
defect formation during symmetry breaking phase
transitions has recently attracted much attention \cite{nature}.  As
originally pointed out by Kibble \cite{kibb} and more recently by Zurek
\cite{zurek}, 
different regions of a system may be unable to correlate
during the quench time which characterizes the transition and, 
as a result, some
parts of space may remain trapped in the original (symmetric)  phase,
giving rise to topological defects. 

Although the Kibble--Zurek mechanism gives a reasonable estimate of the
defect density as a function of the quench time (as confirmed, for
instance, by recent experiments on superfluid Helium 
\cite{helium}), this picture is
essentially phenomenological. It is clear that a full understanding of
the process of defect formation requires a full quantum field
theoretical formulation of the problem. 
There has recently been 
much progress in this direction \cite{Boj}, and here
we give some novel results based on the approach which we are currently
developing. A more systematic account will be presented in
two forthcoming papers \cite{bj1,bj2}.

We are inspired by the work done in the 70's by Umezawa et al. 
\cite{ume2}, who showed how solitons can arise in QFT as result of a
localized (inhomogeneous) condensation of particles. In this picture the
extended objects have an inherently quantum origin. The corresponding
classical soliton solutions are then obtained in the $\hbar \rar 0$ limit. 

Following these ideas we construct a QFT of 
topological defects 
using the Closed--Time--Path (CTP) formalism \cite{CTP}. The
latter is a vital ingredient of our construction, as it has a natural
extension to finite temperature and to non--equilibrium
situations (which is relevant for example in 
realistic phase transitions).

The paper is organized as follows: in Section 2, we give a brief account
of CTP formalism and of some fundamentals of QFT which are essential to
what follows. In Section 3, we consider a simple model, namely $\la
\psi^4$ in $2$ $D$ (i.e.$1+1$), showing how the kink solution can be
constructed at zero and finite temperature. In Section 4, we discuss the
vortex solution for $4D$  complex $\la \psi^4$ theory. 
We conclude in Section 5.

%%%%%%%%%%%%%%%%%%%%%%%%%%%%%%%%%%%%%%%%%%%%%%%%%%%%%%%%%%%%%%%%%%%%%
\vspace{0.3cm}

\noi{\bf  2. The Haag expansion in Closed--Time--Path formalism}

\vspace{0.2cm}

Let us consider the dynamics of an Heisenberg field described by 
\bea\lab{cpt0a} 
\dot{\psi}(x)    =  i[H,\psi(x)] 
\quad, \quad
\dot{\Pi}(x) = i[H, \Pi(x)]\, , 
\eea
where $\Pi$ is the momentum conjugate to $\psi$ and $H$ is the
full (renormalized)  Hamiltonian in the Heisenberg picture.  Assuming that
the Heisenberg and interaction pictures coincide at some time $t_{i}$, we
can write the formal solution of Eqs.(\ref{cpt0a}) as ($``in" $
denotes quantities in the interaction picture) 
\bea\lab{ks0a} 
\psi(x)&=& Z_{\psi}^{1/2}\La^{-1}(t)\;
\psi_{in}(x)\; \La(t) 
%\\ \lab{ks0b}
\quad, \quad 
{\Pi}(x)\,=\,
Z_{\Pi}^{1/2}\La^{-1}(t)\; \Pi_{in}(x)\;\Lambda(t) 
\\ \lab{ks0c} 
\Lambda(t)&=&e^{i(t-t_{i})H_{in}^{0}}\;
e^{-i(t-t_{i})H} \,, \quad
U(t_{2};t_{1}) \, =\,\Lambda(t_{2})\, \Lambda^{-1}(t_{1}) \,=\,
 T\left\{\mbox{exp}\lf[-i\int_{t_{1}}^{t_{2}}\!
d^{4}x \, {\cal H}_{in}^{I}(x)\ri]\right\} 
\eea
where ${\cal H}_{in}^I$ is the interacting Hamiltonian, $T$ is
the time ordering and $Z_{\psi}$, $Z_{\Pi}$ are the wave--function
renormalizations (usually $\Pi \propto \dot{\psi}$, and so $Z_{\psi} =
Z_{\Pi}$). Eqs.(\ref{ks0a}) must be understood in a weak sense:  if not,
we would get the canonical commutator between $\psi$ and $\Pi$
equal to $iZ_{\psi}\de^{3}({\bf x} -{\bf y})$.  This would 
imply $Z_{\psi}=1$,
whilst the K\"allen-Lehmann representation requires $Z_{\psi}<1$.  The
solution of this problem is well known, the Hilbert spaces for $\psi$ and
$\psi_{in}$ are unitarily inequivalent, and the wave function
renormalizations $Z_{\psi}$ and/or 
$Z_{\Pi}$ then ``indicate'' how much
the unitarity is violated \cite{IZ}.
Thus, the choice of the $\psi_{in}$ and of the associated
Hilbert space is not unique: selecting a particular set of $\psi_{in}$ 
corresponds to defining a particular physical situation, i.e.
initial--time
data for the operator equation (\ref{cpt0a}). 
We will use this feature
to construct a QFT which ``contains'' topological defects. 

If derivatives of fields  are not present in   ${\cal L}_{I}$
(and  thus ${\cal H}_{I} = -{\cal L}_{I}$) we can write
\bea\lab{cptdm} \psi(x)  &=&  Z_{\psi}^{1/2}\,U(t_{i};t)\,
\psi_{in}(x)\, U^{-1}(t_{i};t)
\,=\, Z_{\psi}^{1/2}  \,T_{C}  \lf\{
\psi_{in}(x) \exp\lf[ i \int_C d^4x\,  {\cal L}_{in}^{I}(x) \ri]\ri\}
\,.  \eea
Here $C$ denotes a closed--time (Schwinger) contour (see Fig.1),
running from $t_{i}$ to a later time $t_{f}$ and back again.
$T_{C}$ denotes the corresponding time--path ordering symbol. 

\begin{figure}[t]
\vspace{-1cm}
\centerline{\epsfysize=2.5truein\epsfbox{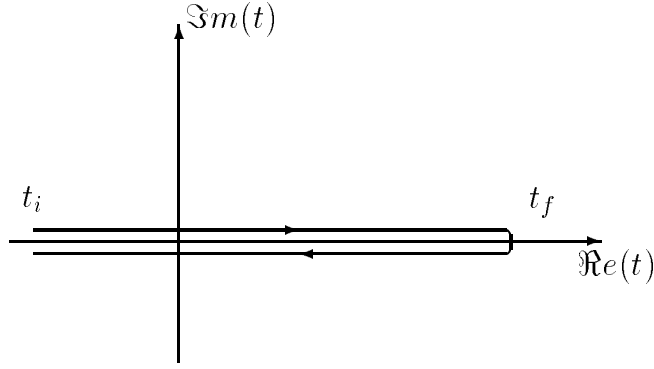}}
\vspace{-0.6cm}
\caption{The Closed Time Path}

\vspace{0.2cm}
\hrule
\end{figure}

In the limit $t_{i} \rar -\infty$, $\psi_{in}$ becomes 
the usual in--(or asymptotic)  field.  Since 
$t_{f}$ is arbitrary,  
it is useful to set $t_{f}= +\infty$. Eq.(\ref{cptdm}) is the so
called Haag expansion, for the Heisenberg field $\psi$ \cite{CTP}. 
 
Generalization of Eq.(\ref{cptdm}) to an arbitrary polynomial 
$P$ of  $\psi$ reads 
\bea\lab{cptdm1} T\lf\{P[\psi]\ri\} =
Z_{\psi}T_{C}\lf\{ P[\psi_{in}]\,
 \mbox{exp}\lf[i\int_{C}d^{4}x \, {\cal L}^{I}_{in}(x)\ri]\ri\}\, .
\eea

\vspace{0.3cm}

%%%%%%%%%%%%%%%%%%%%%%%%%%%%%%%%%%%%%%%%%%%%%%%%%%%%%%%%%%%
\noi{\bf 3. Kinks in two--dimensional $\la \psi^4$ theory}
%%%%%%%%%%%%%%%%%%%%%%%%%%%%%%%%%%%%%%%%%%%%%%%%%%%%%%%%%%%
\vspace{0.2cm}

In this Section we apply the formal considerations developed
above to a specific model, namely the $2D$ $\la \psi^4$ theory both at
zero and finite temperature. Let us consider  the  following Lagrangian
(we adopt the Minkowski signature $(+\, -)$):
\bea\lab{lagr1}  {\cal  L}\,   =\,   \frac{1}{2}(\pa_\mu   \psi)^2   -
\frac{1}{2} \mu^2 \psi^2 - \frac{\la}{4} \psi^4 \, .\eea
The Heisenberg equation of the motion for the field $\psi$ is 
%(to avoid a cumbersome notation we suppress $Z_{\psi}$)
%
\begin{equation}
(\pa^{2} + \mu^{2})\psi(x) = - \la \psi^{3}(x)\, .  \lab{em1}
\end{equation}
For $\mu^{2} < 0$, this 
model admits at a classical level kink solutions. Defining
\bea 
&&\psi(x) =   v + \rho(x) \qquad  {\mbox{with}} \quad v  = \lan
0| \psi(x) |0  \ran \,,\quad
-2\mu^2 =  m^2 \,,\quad -  \mu^2 = \la v^2 \,,\quad g=\sqrt{2
\la}\, , 
\eea
we obtain
\bea\lab{lagr2} 
&& {\cal  L}\,   =\,  \frac{1}{2}(\pa_\mu   \rho)^2    -
\frac{1}{2} m^2 \rho^2 - \frac{g^2}{8} \rho^4 -\frac{1}{2}  m g \rho^3
+ \frac{m^2}{16 \la} 
\\[2mm]
&&(\pa^{2} + m^{2})\rho(x) \,=\,  -\frac{3}{2} m g \rho^{2}(x) -
\frac{1}{2} g^2 \rho^{3}(x) \, .  
\eea
Note that $m^{2}>0$.  The asymptotic field (at $t\rar -\infty$)
now satisfies
\bea 
(\pa^{2} + m^{2})\rho_{in}(x) &=& 0 \, .  
\eea
The 
Haag expansion for the field $\psi$ reads (we suppress the
renormalization factor $Z_{\psi}$)
\bea 
&&\psi(x) = v + T_{C} \lf\{ \rho_{in}(x) \exp\lf[-\frac{i}{2}
\int_C d^2y\,{\cal L}^{I}_{in}(y) \ri]\ri\}\, .  
\eea
where ${\cal L}^{I}_{in} = \frac{g^2}{4} \rho^4_{in} \, +\, m g
\rho^3_{in} $.
We now consider the following (canonical)
transformation:
\bea \lab{cpt3a} 
\rho_{in}(x) \, \rar \, \rho_{in}^{f} = \rho_{in}(x) + f(x) \quad , \quad
(\pa^2 + m^2) f(x) \, =\,0 \, ,  
\eea
with $f$ being a c--number function. It is known \cite{merc} that such a
canonical transformation induces an 
inhomogeneous condensation of $\rho_{in}$ quanta. 

As the shifted field $\rho_{in}^{f}$ fulfils the same asymptotic equations
as the unshifted field $\rho_{in}$ we may  equally well use it as
initial--time data for $\psi$. 
In this case the Haag expansion reads 
\bea 
\psi^f(x) & =& \, v\, +\, T_{C} \lf\{ \rho^f_{in}(x)
\, \mbox{exp}\lf[\frac{-i}{2} \int_C d^2y\,
{\cal L}^{I\, f}_{in}(y)
\ri] \ri\} \, .
\label{41}
\eea
The superscript $f$ in $\psi^f(x)$
 and ${\cal{L}}_{in}^{I\,f}$ indicates that
we work with asymptotic field
$\rho_{in}^{f}$. 
Using the (operatorial) Wick's theorem \cite{IZ}, we rewrite
(\ref{41}) in the following way
\bea \non
\psi^f(x) &=& v\, +\, \lf[\frac{\de}{i\de J(x)}\, +\, f(x) \ri]\,
\mbox{exp}\lf\{\frac{-i}{2} \int_C d^2y\, {\cal L}^{I\,
f}_{in}\lf[\frac{\de}{i\de J} , y\ri]
\ri\}  
\\[2mm]
&& \times \,
:\exp \lf[ i \int_C d^2y\, J(y)\rho_{in}(y)\ri]: \,\lf.  \exp
\lf\{-\frac{1}{2}\int_C d^2y d^2z\,J(y)\De_C(y;z)J(z)\ri\} 
\ri|_{J=0}\,
, \lab{cpt72} 
\eea
 where $\De_C (x;y)=\lan T_C[\rho_{in}(x)\rho_{in}(y)] \ran_{}^{}$ ($\lan
\ldots \ran\, \equiv\lan 0|\ldots| 0\ran $ where $|0\ran$ 
is the vacuum for the $\rho_{in}$ field).   
We now take the vacuum expectation value (vev) 
of Eq.(\ref{cpt72}), use the 
relation $ F\lf[ \frac{\de}{i\de J} \ri] G[J] = \lf.  G\lf[
\frac{\de}{i\de K} \ri] F[K] e^{i K J}\ri|_{K=0}^{}
$, perform the change of variables $K \rar K\, +\,f$ and set to zero the
term with $J$ (there are no 
derivatives with respect to it). The result is \cite{bj1}
\bea
\lan \psi^f(x) \ran = v\,+\, \left.\mbox{exp}[\hbar a]
\,K(x)\, \mbox{exp}\left[\frac{1}{\hbar} b\right] \ri|_{K=f} \, ,
\lab{cpt10}
\eea
where we have reintroduced $\hbar$ and defined 
\bea
a \equiv -\frac{1}{2}\int_{C}d^{2}zd^{2}y \,\De_{C}(z,y)\,
\frac{\de^{2}}{\de K(z) \de K(y)} \quad, \quad 
b \equiv-\frac{i}{2}\int_{C}d^2z\,\lf [
\frac{g^2}{4}\,K^4(z)\,+\, m g\,K^3(z)\ri]\, .
\eea
\noi After some manipulations we arrive at the following form for the
order parameter\cite{bj1}  
\bea\lab{ordpar1} 
&&\lan \psi^f(x)\ran = v \, +\,\lf.  C[K](x)\ri|_{K=f} \, +\, \lf.
D[K](x)\ri|_{K=f} \, ,\\[2mm] 
&& C[K](x) = - \int_{C}d^{2}y \Delta_{C}(x,y) \frac{\delta}{\delta K(y)}
\, \mbox{exp}[\hbar a]\, 
\mbox{exp}\left[\frac{1}{\hbar} b\ri] \,,
\\[2mm] 
&& D[K](x) = K(x)\, \mbox{exp}[\hbar a]\,\mbox{exp}\left[\frac{1}{\hbar}
b\ri]  
\, .  \eea 
We remark that the order parameter Eq.(\ref{ordpar1}) still contains all
quantum corrections.

\vspace{0.2cm}

\noi {\bf Classical kinks at zero temperature}

Let us now deal with the Born, or classical, approximation of
(\ref{ordpar1}).  For this purpose we keep 
only finite parts in $C$ and $D$ terms in the $\hbar \rar 0$ limit. We
then get:  
\bea \lab{CK}
C(\hbar \rar 0)|_{finite}  &=& 
\mbox{Res}_{\hbar =0} \lf[C(\hbar) \ri] \, =\,
- i\int_{-\infty}^\infty d^{2}y\, G_{R}(x,y) \, \frac{\de}{\de
K(y)}\sum_{n=0}^{\infty} \frac{1}{n!(n+1)!}  \, a^{n}b^{n+1} \, ,
\\ \lab{DK} 
D(\hbar \rar 0)|_{finite} &=&
\mbox{Res}_{\footnotesize \hbar = 0} \lf[ D(\hbar)\ri]\,=\, 
K(x)\,\sum_{n=0}^{\infty}\,\frac{1}{(n!)^{2}}\,a^{n}b^{n}\, . 
\eea
where $iG_{R}(x,y)
=\theta(x_{0}-y_{0})\, \Delta(x,y)$ is the  retarded 
Green's function of the free theory and 
$\De(x;y)\equiv \lan [\rho_{in}(x),\rho_{in}(y)] \ran $ 
is the Pauli--Jordan function:
\begin{equation}\lab{pjordan}
\De(x_{0},x_{1};0,0) =  \int
\frac{d^{2}k}{(2\pi)}\, \de(k^2 - m^2)\,\varepsilon(k_{0})\,
e^{-ikx}\non = -\frac{i}{2} \,\theta(x_{0} - |x_{1}|)\,
J_{0}(m\, 
\sqrt{x_{0}^2- x_{1}^2})\, .
\end{equation}
Finally we may write \cite{bj1}  
\bea
 \lan \psi_{0}^{f}(x)\ran &= & 
\,v\,+ \, \sum_{n=1}^\infty Q_n[K](x) \Big|_{K=f}
\, , 
\eea
where $ Q_1 = f $  and ($i,j,k =1,2
\ldots$ and $n>1$. An empty sum is zero)
\bea\lab{45}
Q_n(x) &=&
-\int_{-\infty}^{\infty}\!d^{2}y\, G_{R}(x,y)\,\lf[ \frac{3}{2}m g
\!\!\sum_{i+j=n} \!\! Q_i(y)\, Q_j(y) \, +\, \frac{1}{2} g^2
\!\!\!\sum_{i+j+k=n} \!\!\! Q_i(y)\, Q_j(y)\, Q_k(y) \ri] \, ,
 \eea
This recurrence 
relation belongs to the class of the so called functional equations
of Cauchy--Marley's 
type\cite{Bell,Mar} whose fundamental solution
cannot be expressed (apart from a very 
narrow class of kernels) in terms of elementary functions. This means
that we cannot hope to resolve 
(\ref{45}) in terms of general $K$ (or $f$).  We can however obtain
analytical solutions of the 
classical version of Eq.(\ref{em1})  - the analytical kinks. These may
be obtained if we realize that the 
convolution of the $2D$ retarded Green's function $G_{R}(x)$ with an
exponential is proportional to the very 
same exponential. 
Thus, if $Q_{1}(x)  = f(x)$ is an 
exponential, Fourier non--transformable, solution of Eq.(\ref{cpt3a}),
we have $Q_{n}(x) \propto (Q_{1}(x))^{n}$ and 
\bea Q_{n}(x) =
A_{n}f^{n}(x) = A_{n}\, e^{\pm m n\gamma(x_{1} -x_{0}u)}\, ,
\eea
where $\gamma = (1-u^2)^{-\frac{1}{2}}$.
Plugging this form into (\ref{45}) we arrive at
the following equation 
\bea A_{n}=
\frac{1}{(n^{2}-1)}\left\{\frac{3}{2v}\sum_{i+j=n} A_{i}A_{j} + \frac{1}{2
v^{2}}\sum_{i+j+k=n} A_{i}A_{j}A_{k}\ri\}\, .  \lab{An} 
\eea 
A solution of 
(\ref{An}) is $A_n\, =\, 2 v \, \lf(\frac{s}{2v}\ri)^n$, with
$s$ being a real constant. The order parameter is
\bea \lab{sol1} \lan \psi_{0}^{f}(x)\ran \, &=& \, v +
2v\sum_{n=1}^{\infty}\lf(\frac{s f(x)}{2v}\ri)^{n} \, . 
\eea
Thus, provided $f(x)$ is an exponential solution of the linear
equation (\ref{cpt3a}), the solution (\ref{sol1}) fulfils the (classical) 
Euler--Lagrange equation of motion:  
\bea\lab{euler} (\pa^{2} +
\mu^{2})\lan \psi_{0}^{f}(x)\ran \, =\, - \la \,\lan
\psi_{0}^{f}(x)\ran^{3} \, , \eea 
which is nothing but the vev of Eq.(\ref{em1}) in the Born approximation. 
For instance, if we choose  $f(x) = 
e^{- m \gamma(x_{1} -x_{0}u )}$ with $s=-2v e^{m \gamma
a}$, we obtain the standard kink solution\cite{Da,Go,Po} 
\bea
\lan \psi_{0}^{f}(x) \ran \,= \,v \;
\mbox{th}\lf[\frac{m}{2}\gamma \Big((x_{1}-a) -x_{0}u\Big)\ri]\, , 
\eea
describing a constantly moving kink of a permanent profile with
a center localized at $a+ux_{0}$.  

\vspace{0.2cm}

\noi {\bf Classical kinks at finite temperature}

Let us now concentrate on the finite temperature case. It is important
to understand what happens with the 
topological defects if the system is immersed in a heat bath at
temperature $T$. 

The crucial observation at finite temperature is that the
operatorial Wick's theorem still holds \cite{DS} and thus
Eq.(\ref{cpt72}) retains its validity provided ($\ll
\ldots\gg $ denotes thermal average)
\bea
\Delta_{C}(x,y)\, =\, \langle 0| T_{C}(\rho_{in}(x) \rho_{in}(y)) | 0
\rangle \, \rightarrow \, \Delta_{C}(x,y;T) \,=\, \ll T_{C}(\rho_{in}(x)
\rho_{in}(y)) \gg \, ,
\eea
together with $ :\ldots :\; \rightarrow \; N(\ldots )$. 
The thermal normal ordering $N(\ldots)$ is defined is
such a 
way\cite{DS} that $\ll N(\ldots ) \gg = 0 $. This is of
a great importance as all the previous formal considerations 
go through also for finite $T$.

At finite temperature the question of $\hbar$ appearance is more
delicate than in the zero--temperature 
case. The whole complication is hidden in the thermal propagator
$\Delta_{C}(x,y;T)$. To understand this, 
let us make $\hbar$ explicit. The free thermal
propagator in spectral or Mills's 
representation\cite{Mills,LB} then reads
\bea\lab{prop2}
i\Delta_{C}(x,y;T) &=& \hbar\int\frac{d^{2}k}{(2\pi)^{2}} \,
e^{-ik(x-y)}\,\rho(k) [\theta_{C}(x_{0}-y_{0}) + f_{b}(\hbar
k_{0}/T) ]\nonumber \\ 
&=& i\Delta_{C}(x,y;0) + i\Delta^{T}_{C}(x,y)
\, ,
\eea
where the spectral density $\rho(k) =
(2\pi)\varepsilon(k_{0})\, \delta(k^{2}-m^{2})$ with
$\varepsilon(k_{0}) = \theta(k_{0}) - \theta(-k_{0})$. The contour
step function $\theta_{C}(x_{0}-y_{0})$ is $1$ if $y_{0}$ precedes
$x_{0}$ along the contour $C$. The Bose--Einstein distribution 
$f_{b}(x)= \left(e^{x}-1\right)^{-1}$. To calculate $Res_{\hbar
\rightarrow 0}$ we need to perform a Laurent expansion of
$f_{b}$ around $\hbar$, i.e.
\bea
f_{b} + \frac{1}{2} = \frac{T}{\hbar k_{0}} + \frac{1}{12} \frac{\hbar
k_{0}}{T} + \ldots \, , 
\eea
which converges for $\hbar |k_{0}| < 2\pi T$. Taking a
regulator $\Lambda \sim \frac{\hbar}{T}$ in 
$k_{0}$ integration, i.e considering only {\em soft modes} we
get \cite{bj2}  (see also ref.\cite{vitman})
\bea
\ll \psi^{f}(x) \gg = v(T) \mbox{th}\left[ \frac{m(T)}{2} \gamma
\Big( (x_{1} - a) - x_{0}u\Big)\right]\, , 
\eea
where $v^{2}(T) = \frac{1}{\lambda}\lf(|\mu^{2}| - 3\lambda \ll
:\rho^{2}_{in}: \gg \ri)$ and $m(T) = 
\sqrt{2\lambda}v(T)$. Thus, at a ``critical" temperature defined by
equation: $ |\mu^{2}| = 3\lambda 
\ll: \rho^{2}_{in}: \gg$, the kink disappears. 

It is an interesting
question to ask whether for higher dimensional systems, this temperature
is related  to the critical temperature of the system.

%\vspace{3mm}
\newpage 

%%%%%%%%%%%%%%%%%%%%%%%%%%%%%%%%%%%%%%%%%%%%%%%%%%%%%%%%%%%%%%
\noi{\bf 4. Vortices in four--dimensional $\la \psi^4$ theory}
%%%%%%%%%%%%%%%%%%%%%%%%%%%%%%%%%%%%%%%%%%%%%%%%%%%%%%%%%%%%%%

\vspace{0.2cm}

In this Section we sketch the treatment of the $4D $ complex $ \la
\psi^4$ theory in presence of vortices. For a more thorough
discussion see ref.\cite{bj2}. We show how the vortex solution can
be obtained by ``shifting" both the fields of the massless (Goldstone)
mode  and  of
the  massive (unstable) mode: the appearance of a topological charge
is controlled by the shift of the  Goldstone field only.
We consider the Ginzburg--Landau type  Lagrangian for  a
charged scalar field: 
\bea\label{vor0}  
{\cal L}\, =\,  \pa_\mu  \psi^\dag \pa^\mu\psi - \mu^2
 \psi^\dag\psi - \frac{\la}{4} |\psi^\dag\psi|^2 \, .
\eea 
The (unrenormalized) equations of motion for the Heisenberg operator
$\psi$ read 
\bea\label{vor1} 
\lf(\Box \, +\, \mu^2  \,+ \, \frac{\la}{2} |\psi(x)|^2
\ri)\psi(x)\, =\,0 \, . 
\eea 
We assume symmetry breaking,
i.e.  $\mu^2<  0$  and  $\lan  \psi  \ran  =  v=\sqrt{-2\mu^2/\la}$.  We
parametrize the  field as   $\psi(x)\equiv (\rho(x)   +v) e^{i\chi(x)}$,
where both  the  fields $\rho$ and  $\chi$  are hermitian and  have zero
vev. We
also put $g=\sqrt{\la}$ and $m^2=\la v^2  > 0$.  The equations of motion
for  $\rho$ and $\chi$ read 
\bea 
\label{vor2a}
&&\lf[\Box -   (\pa_\mu \chi)^2  \,+\,m^2\ri]\rho +  \frac{3}{2}m\,  g\,
\rho^2 + \frac{1}{2} g\,\rho^3
\,=\, v \,(\pa_\mu \chi)^2 
\\ [2mm] 
\label{vor2b} 
&&\pa_\mu \lf[ (\rho + v )^2\pa^\mu \chi\ri]\, =\, 0 \, . 
\eea 
We now choose  as the relevant asymptotic  fields the ones  described by
the equations
\bea 
\label{vor2c}
&&\lf[\Box - (\pa_\mu  \chi_{in})^2 \,+\,m^2\ri](\rho_{in} \, + \, v)
\,=\,  0 
\\ [2mm] \label{vor2d}   
&&\pa_\mu \lf[(\rho_{in} +v)^2 \,  \pa^\mu
\chi_{in}\ri]\, =\, 0 \, ,\eea 
i.e. $\lf(\Box \, +\, m^2 \ri)\psi_{in}\, =\, 0 $\, ,
where we have used  $\psi_{in}\equiv (\rho_{in}+v)\, e^{i\chi_{in}}$.   
This means that the interaction Lagrangian density is
\bea 
\lab{Lint} 
{\cal L}^I& =& (m^2- \mu^2) \psi^\dag\psi \,  -\, 
\frac{\la}{4} |\psi^\dag\psi|^2
\, =\, 2 m^2 v \rho\,  - \,g m \rho^3
\,- \,\frac{g^2}{4} \rho^4 \, +\, \frac{5}{4}m^2v^2\, .  
\eea 

The corresponding Haag expansion for the Heisenberg field 
operator $\psi$ is then
\bea \label{vor4} 
\psi(x)\equiv (\rho(x)+v)  e^{i\chi(x)} \,  
 =\, T_C\lf\{(\rho_{in}(x) + v) e^{i\chi_{in}(x)} \exp\lf[-i\int_C 
d^4y {\cal L}^I_{in}(y)\ri] \ri\} \, .  
\eea 
As before, the  solutions  of the asymptotic equations
(\ref{vor2c}) and (\ref{vor2d})  are not unique. Indeed, we  may define
the following shifted fields: 
\bea\lab{shift1}  
&&\rho_{in}(x)\rar\rho^f_{in}(x)   \,=\,\rho_{in}(x)
\,+\,f(x) 
\\ [2mm] \lab{shift2} 
&&\chi_{in}(x)\rar\, \chi^g_{in}(x)=\,
\chi_{in}(x) \,+\,g(x)\, , 
\eea 
satisfying the same equations as the ones for the unshifted fields. The 
c--number functions  $h\equiv f+v$ 
and $g$ solve the  coupled
equations:
\bea \label{eqfg1}
&&\lf[\Box - \lan(\pa_\mu \chi_{in}(x))^2\ran  - (\pa_\mu g(x))^2 + m^2
\ri]h(x)\,=\,m^2 v    
\\[2mm] \label{eqfg2}     
&&\pa_\mu \lf[ \lf(
\lan\rho_{in}^2 \ran + h^2(x) \ri)\pa^\mu g(x)\ri]\, =\, 0 \, . 
\eea 

The Haag expansion for  $\psi$
in terms of the new asymptotic fields reads 
\bea  \label{vor8}                  
&&\psi^{f,g}(x)     \,=\,
T_C\lf\{(\rho^f_{in}(x)+v)e^{i\chi^g_{in}(x)} 
\exp\lf[-i\int_C d^4y {\cal L}^{I\,f}_{in}(y)\ri] \ri\}, 
\eea 
By  construction the   field
$\psi^{f,g}(x)$  satisfies  the same Heisenberg   equations as the field
$\psi(x)$.   A   particular   choice  of    solutions  of
Eqs.(\ref{eqfg1}), (\ref{eqfg2}) can lead  to    the  description of   a
particular physical  situation, e.g. of a system with topological
defects. Observe that in (\ref{vor8}) the whole dependence on $g$
factorises out of $T_{C}$. 

Let us now consider the
vev of the
Heisenberg operator (\ref{vor8}), i.e. the order parameter: 
\bea \label{vor8b}
\lan \psi^{f,g}(x) \ran \equiv e^{i g(x)} \, F_f(x)
%\\ &&
\, =\,e^{i g(x)}  \lan 0|  T_C\lf\{(\rho^f_{in}(x) + v)  
e^{i\chi_{in}(x)} exp\lf[-i\int_C d^4y {\cal L}^{I\,f}_{in}(y)\ri] 
\ri\} |0 \ran . 
\eea 
In the classical approximation, the order parameter satisfies
\bea \label{vor9a}
&&\lf[\Box - (\pa_\mu g(x))^2 +m^2 +\la F_f^2(x) \ri] F_f(x)\,=\,0 
\\  \label{vor9b} 
&&\pa_\mu \lf[ F_f^2(x) \pa^\mu g(x)\ri]\, =\, 0 \, .
\eea 
This is nothing but
the well known vortex equations
\cite{kleinert}. A particular solution -- a static
vortex along the  third axis --  is obtained taking $F_f(x)$ time
independent  with a radial dependence only. 
Eq.(\ref{vor9b}) then reduces to the Laplace  equation having the 
polar angle as solution:
\bea\label{vor10}   
g(x)\,=\, n\,\te(x)\,=\, n\,
\tan^{-1}\lf(\frac{x_2}{x_1}\ri)  \, ,
\eea 
where   the integer $n$ guarantees the   single valuedness  of the order
parameter $\lan \psi^{f,g}(x)\ran $. $F_f(r)$ then fulfils 
 the following (static) equation: 
\bea\label{vor11}  
\lf[\pa_r^2 +\frac{1}{r}\pa_r  -\frac{n^2}{r^2} +m^2
\ri] F_f(r)\,=\, \la F_f^3(r) \, . \eea 

At this point we can use the fact that the function $g$
appearing in the
vortex equations (\ref{vor9a}), (\ref{vor9b}) and in the equations for
the shift functions (\ref{eqfg1}), (\ref{eqfg2}) {\em is the same}:
we pick up the solution of the static vortex equation and plug it into the
shift equations(\ref{eqfg1}), (\ref{eqfg2}) to 
determine\footnote{In doing this, we use the
renormalized equations, thus the terms with the
vev will drop. The notation is kept in (\ref{eqfg1}), (\ref{eqfg2})
since 
in general these terms are present (eg. at finite temperature as thermal
averages). } 
the other shift function, $f$.
Let us denote this solution by ${\tilde f}$. 
Then the expression for the Heisenberg operator  in  the presence of a
static vortex at $x=0$ will be 
\bea \label{vor15}
 \psi^{vort.}(x) & =& T_C\lf\{\lf(\rho_{in}(x) + {\ti f}(r(x))+
v\ri) \, e^{i\chi_{in}(x)+i n \te(x)} 
\exp\lf[-i\int_C d^4y  {\cal L}^{I\,{\ti f}}_{in}(y)\ri] \ri\}\, .
\eea 
This expression can be used as a starting point for further
analysis \cite{bj2}. Here we only consider the simplest
approximation, i.e. we completely neglect the radial dependence by
setting $f=0$. Then the following relation holds:
\bea
\psi^{\te}(x) & \simeq & e^{i n \te(x)} \, \psi(x) \, ,
\eea
from which also follows the relation between Green's functions with and
without vortices 
($x$ and $y$ are far from the vortex core): 
\bea\label{bo7} 
G^\te(x,y)&\equiv& \lan 0|T( \psi^\te(x) \psi^{\te \dag}(y)) |
0 \ran \, 
 \simeq \,e^{in \lf[\te(x) - \te(y)\ri] } \, G(x-y) \, .
\eea 

\noi It is interesting to note that a similar relation can be derived for
a systems of (charged) particles in a Bohm--Aharonov potential. In
this case, one has \cite{schul}:
\bea\label{bo9}  
G_{\bf A}(x,y)  \, =\,e^{-i e  \lf[\Om(x)  - \Om(y)\ri] } \,G_0(x,y)\, ,
\eea 
where the Bohm--Aharonov potential is that induced by a magnetic flux = $n$
times the elementary magnetic flux:
\bea
{\bf A} \,=\, \nabla \Om \quad, \quad
\oint d{\bf s} \cdot {\bf A} \,=\, - 2 \pi n /e \, .
\eea

\noindent In the case of a vortex line of strength $n$ we also have,
beside
Eq.(\ref{bo7}):
\bea
{\bf J} \,=\, n \nabla \te \quad, \quad
\oint d{\bf s} \cdot {\bf J}\,=\, 2 \pi n \, .
\eea
\noindent Thus the ``duality" correspondence reads
$n\te(x)\, \lrar \, -e\Om(x)$
 and 
${\bf J}\, \lrar \, - e {\bf A}$.

\vspace{0.5cm}

%%%%%%%%%%%%%%%%%%%%%%%%%%%%%%%%%%%%%%%%%%%%%%%%
\noi{\bf 5. Conclusions}

\vspace{0.2cm}
We have developed a field theoretical (operator) formalism suitable
for the description of quantum systems containing (topological)
defects. 
The use of the Closed--Time--Path formulation is crucial in our
approach, since it allows to treat systems both at zero temperature
or in thermal equilibrium as well as systems out of equilibrium.

We have applied our method to $\la \psi^4$ theory in $2D$ and $4D$
cases. In $2D$ case, kink solutions 
were studied at zero and at finite temperature. In $4D$ case, vortex
solution was shown to arise from 
a inhomogeneous condensation of Goldstone modes; an analogy with
Bohm--Aharonov effect was discussed.

\vspace{0.5cm}

\noi{\bf Acknowledgements}

\vspace{0.2cm}

We would like thank T.Kibble and G.Vitiello 
for discussions and encouragement.
We also thank the organizers of the School, A.Zichichi,
G.'t Hooft and  G.Veneziano, and  M.Gourdin 
for the support received by P.J. as  a Scholarship
and for the invitation to contribute to these Proceedings. 
This work has been partially supported by MURST, INFN and ESF.

\end{document}